\documentclass{article}%
\usepackage{amsmath}
\usepackage{amsfonts}
\usepackage{amssymb}
\usepackage{graphicx}%
\providecommand{\U}[1]{\protect\rule{.1in}{.1in}}

\begin{document}

\title{Riemann-Cartan Connection and its Decomposition. One More Assessment of
\textquotedblleft ECE Theory\textquotedblright}
\author{J. Fernando T. Giglio$^{(1)}$ and Waldyr A. Rodrigues Jr.$^{(2)}$\\$^{(1)}$ FEA-CEUNSP. 13320-902 - Salto, SP Brazil.\\$^{(2)}$ Institute of Mathematics Statistics and Scientific Computation\\IMECC-UNICAMP \\13083950 Campinas, SP Brazil\\walrod@ime.unicamp.br or walrod@mpc.com.br}
\maketitle

\begin{abstract}
In this short pedagogical note we clarify some subtleties concerning the
symmetries of the \emph{coefficients} of a Riemann-Cartan \textit{connection}
and the symmetries of the coefficients of the \textit{contorsion} tensor that
has been a source of some confusion in the literature, in particular in a so
called `ECE theory'. We show in details that the coefficients of the
contorsion tensor of a Riemann-Cartan connection has a symmetric part and an
antisymmetric part, the symmetric part defining the \emph{strain tensor }of
the connection. Moreover,\ the contorsion tensor has also a \textit{bastard}
anti-symmetry when written with all its indices in the\ `covariant' positions.

\end{abstract}

\section{Some Preliminaries}

Let $M$ be a 4-dimensional\ Hausdorff, paracompact and locally compact
manifold admitting a Lorentzian metric tensor $\boldsymbol{g}\in\sec T_{0}%
^{2}M$. Let us suppose that $M$ is also spacetime orientable by a global
$4$-form field $\tau_{\boldsymbol{g}}\in\sec%
{\textstyle\bigwedge\nolimits^{4}}
T^{\ast}M$ and also time orientable\footnote{See, e.g., \cite{sawu} for
details.
\par
{}} by the relation $\uparrow$ and let be $\mathring{D}$ the Levi-Civita
connection of $\boldsymbol{g}$. Under these conditions we call the pentuple
$\langle M,\boldsymbol{g},\mathring{D},\tau_{\boldsymbol{g}},\uparrow\rangle$
a \textit{Lorentzian spacetime}. The curvature tensor of $\mathring{D}$ will
be denoted in what follows by $\boldsymbol{\mathring{R}}$.\medskip

Let $D$ be a general Riemann-Cartan connection on $M$, $i.e$.,
$D\boldsymbol{g}=0$. In general the Riemann (curvature) tensor $\boldsymbol{R}%
$ and the torsion tensor $\boldsymbol{\Theta}$ of $D$ are non null. Under the
conditions of orientability and time orientability the pentuple $\langle
M,\boldsymbol{g},D,\tau_{\boldsymbol{g}},\uparrow\rangle$ is said to be a
Riemann-Cartan spacetime.\medskip

By definition a manifold equipped with a given connection is said to be\emph{
flat} if the Riemann (curvature) tensor of that connection is null.\medskip

Let $\langle x^{\mu}\rangle$ and $\langle x^{\prime\mu}\rangle$ be
respectively coordinate functions for $U\subset M$ and $U^{\prime}\subset M$
such that $U\cap U^{\prime}\neq\varnothing$.

Moreover, let $\langle\boldsymbol{e}_{\mu}=\partial/\partial x^{\mu}\rangle$
and $\langle\boldsymbol{e}_{\mu}^{\prime}=\partial/\partial x^{\prime\mu
}\rangle$ be respectively basis of $TU$ and $TU^{\prime}$ $(\mu=0,1,2,3)$ and
$\langle\vartheta^{\mu}=dx^{\mu}\rangle$ and $\langle\vartheta^{\prime\mu
}=dx^{\prime\mu}\rangle$ the corresponding dual basis \ i.e., basis for
$T^{\ast}U$ and $T^{\ast}U^{\prime}$. We also introduce the reciprocal basis
$\langle\boldsymbol{e}^{\mu}\rangle$\ of $\langle\boldsymbol{e}_{\mu}\rangle$
and $\langle\boldsymbol{e}^{\prime\mu}\rangle$\ of $\langle\boldsymbol{e}%
_{\mu}^{\prime}\rangle$ for $TU$ and $TU^{\prime}$ and the reciprocal basis
$\langle\vartheta_{\mu}\rangle$ of $\langle\vartheta^{\mu}\rangle$ and
$\langle\vartheta_{\mu}^{\prime}\rangle$ of $\langle\vartheta^{\prime\mu
}\rangle$ for $T^{\ast}U$ and $T^{\ast}U^{\prime}$, such that%
\begin{align}
\boldsymbol{g}  &  =g_{\mu\nu}\vartheta^{\mu}\otimes\vartheta^{\nu}=g^{\mu\nu
}\vartheta_{\mu}\otimes\vartheta_{\nu},\text{ \ \ }g^{\mu\alpha}g_{\alpha\nu
}=\delta_{\nu}^{\mu},\nonumber\\
\boldsymbol{e}^{\mu}  &  =g^{\mu\nu}\boldsymbol{e}_{\nu},\text{ \ \ }%
\vartheta_{\mu}=g_{\mu\nu}\vartheta^{\nu},\text{etc.} \label{rec}%
\end{align}
Moreover we introduce as metric for the cotangent bundle the object $g\in\sec
T_{2}^{0}M,$%
\[
g=g^{\mu\nu}\boldsymbol{e}_{\mu}\otimes\boldsymbol{e}_{\nu}=g_{\mu\nu
}\boldsymbol{e}^{\mu}\otimes\boldsymbol{e}^{\nu}%
\]
\noindent and define the scalar product of arbitrary of arbitrary vector
vector fields $\mathbf{V},\mathbf{W}\in\sec TM$ $\ $and arbitrary $1$-form
fields $\boldsymbol{X}$ $,\boldsymbol{Y}$ $\in\sec%
{\textstyle\bigwedge\nolimits^{1}}
T^{\ast}M$ by
\begin{equation}
\mathbf{V}\cdot\mathbf{W}=\boldsymbol{g}(\mathbf{V},\mathbf{W}),\text{
\ \ }\boldsymbol{X}\cdot\boldsymbol{Y}=g(\boldsymbol{X},\boldsymbol{Y}).
\label{SP}%
\end{equation}

We write%
\begin{align}
D_{\boldsymbol{e}_{\mu}}\vartheta^{\nu}  &  :=-\Gamma_{\mu\alpha\cdot}%
^{\cdot\cdot\nu}\vartheta^{\alpha},\text{ \ \ }D_{\boldsymbol{e}_{\mu}%
^{\prime}}\vartheta^{\prime\nu}:=-\Gamma_{\mu\alpha\cdot}^{\prime\cdot\cdot
\nu}\vartheta^{\prime\alpha},\nonumber\\
\mathring{D}_{\boldsymbol{e}_{\mu}}\vartheta^{\nu}  &  :=-\mathring{\Gamma
}_{\mu\alpha\cdot}^{\cdot\cdot\nu}\vartheta^{\alpha},\text{ \ \ }\mathring
{D}_{\boldsymbol{e}_{\mu}^{\prime}}\vartheta^{\prime\nu}:=-\mathring{\Gamma
}_{\mu\alpha\cdot}^{\prime\cdot\cdot\nu}\vartheta^{\prime\alpha}. \label{Ds}%
\end{align}

\section{$\Gamma_{\rho\sigma\cdot}^{\cdot\cdot\mu}$ is not in General
Antisymmetric in the Lower Indices}

As it is well known, given an arbitrary connection $D$, the relation between
$\Gamma_{\iota\kappa\cdot}^{\prime\cdot\cdot\lambda}$ and $\Gamma_{\rho
\sigma\cdot}^{\cdot\cdot\mu}$ (dubbed transformation law for the connection
coefficients) is%
\begin{equation}
\Gamma_{\iota\kappa\cdot}^{\prime\cdot\cdot\lambda}=\frac{\partial
x^{\prime\lambda}}{\partial x^{\mu}}\frac{\partial x^{\rho}}{\partial
x^{\prime\iota}}\frac{\partial x^{\sigma}}{\partial x^{\prime\kappa}}%
\Gamma_{\rho\sigma\cdot}^{\cdot\cdot\mu}+\frac{\partial x^{\prime\lambda}%
}{\partial x^{\mu}}\frac{\partial^{2}x^{\mu}}{\partial x^{\prime\iota}\partial
x^{\prime\kappa}} \label{d-d'}%
\end{equation}
From Eq.(\ref{d-d'}) we see that even if happens that $\Gamma_{\rho\sigma
\cdot}^{\cdot\cdot\mu}$ is \textit{antisymmetric }in a given coordinate basis,
i.e., $\Gamma_{\rho\sigma\cdot}^{\cdot\cdot\mu}=-$ $\Gamma_{\sigma\rho\cdot
}^{\cdot\cdot\mu},$ in general it will be $\boldsymbol{not}$ antisymmetric in
another coordinate chart since the term $\frac{\partial x^{\prime\lambda}%
}{\partial x^{\mu}}\frac{\partial^{2}x^{\mu}}{\partial x^{\prime\iota}\partial
x^{\prime\kappa}}$ is symmetric in the lower indices. This immediately
contradicts the main claim of the so called \textquotedblleft ECE unified
field theory\textquotedblright\ \footnote{See the criticisms to ECE theory in
the list of references. Particularly, see \cite{5} where Bruhn, pedagogicaly
identifies that the mistake of the author of ECE papers regarding his
statement that\ for any general connection its coefficients in any coordinate
basis must be anti-symmetric\ is simply due to the fact that he did not know
(until today) that a general real $n\times n$ matrix \ can be decomposed in a
symmetric matrix plus an anti-symmetric one.}, where it is stated that the
connections coefficients of a Riemann-Cartan connection are \textit{always} antisymmetric.

\section{Relation between $\Gamma_{\mu\nu\cdot}^{\cdot\cdot\lambda}$ and
$\mathring{\Gamma}_{\mu\nu\cdot}^{\cdot\cdot\lambda}$}

We shall prove that :%

\begin{equation}
\Gamma_{\mu\nu\cdot}^{\cdot\cdot\lambda}=\mathring{\Gamma}_{\mu\nu\cdot
}^{\cdot\cdot\lambda}+K_{\mu\nu\cdot}^{\cdot\cdot\lambda}\label{3a}%
\end{equation}
where\footnote{Note that this formula differs by a factor of $1/2$ and signal
from the one in \cite{4}.}
\begin{align}
K_{\mu\nu\cdot}^{\cdot\cdot\beta} &  :=\frac{1}{2}(T_{\mu\nu\cdot}^{\cdot
\cdot\beta}+S_{\mu\nu\cdot}^{\cdot\cdot\beta})\nonumber\\
&  =\frac{1}{2}g^{\lambda\beta}g_{\lambda\alpha}T_{\mu\nu\cdot}^{\cdot
\cdot\alpha}-\frac{1}{2}g^{\beta\lambda}g_{\nu\rho}T_{\mu\lambda\cdot}%
^{\cdot\cdot\rho}-\frac{1}{2}g^{\beta\lambda}g_{\mu\alpha}T_{\nu\lambda\cdot
}^{\cdot\cdot\alpha}\text{ }\label{3b}\\
&  =\frac{1}{2}(T_{\mu\nu\cdot}^{\cdot\cdot\hspace{0.01in}\beta}-T_{\nu
\cdot\mu}^{\cdot\beta\cdot}+T_{\cdot\mu\nu}^{\beta\cdot\cdot}).\nonumber
\end{align}
and%

\begin{align}
T_{\mu\nu\cdot}^{\cdot\cdot\lambda}  &  =\Gamma_{\mu\nu\cdot}^{\cdot
\cdot\lambda}-\Gamma_{\nu\mu\cdot}^{\cdot\cdot\lambda}=-T_{\nu\mu\cdot}%
^{\cdot\cdot\lambda},\label{4}\\
S_{\mu\nu\cdot}^{\cdot\cdot\lambda}  &  =-g^{\lambda\sigma}(g_{\nu\alpha
}T_{\mu\sigma\cdot}^{\cdot\cdot\alpha}+g_{\mu\alpha}T_{\nu\sigma\cdot}%
^{\cdot\cdot\alpha})=S_{\nu\mu\cdot}^{\cdot\cdot\lambda}\text{ }. \label{5}%
\end{align}

Before presenting the proof (see also \cite{1,2}) of the above equations we
recall that the $T_{\mu\nu\cdot}^{\cdot\cdot\lambda}$ are the components of
the so called \textit{torsion tensor}$\ $%
\begin{equation}
\boldsymbol{\Theta}=\frac{1}{2}T_{\mu\nu\cdot}^{\cdot\cdot\lambda}%
\vartheta^{\mu}\wedge\vartheta^{\nu}\otimes\boldsymbol{e}_{\lambda}\in\sec%
{\textstyle\bigwedge\nolimits^{2}}
T^{\ast}M\otimes TM, \label{7}%
\end{equation}
Also, the $K_{\mu\nu\cdot}^{\cdot\cdot\lambda}$ are the components of an
object that\ (since Schouten \cite{3a}) is called the con\textit{torsion
tensor}
\begin{equation}
\boldsymbol{K}=K_{\mu\nu\cdot}^{\cdot\cdot\lambda}\vartheta^{\mu}%
\otimes\vartheta^{\nu}\otimes\boldsymbol{e}_{\lambda}=K_{\mu\nu\lambda}%
^{\cdot\cdot\cdot}\vartheta^{\mu}\otimes\vartheta^{\nu}\otimes\boldsymbol{e}%
^{\lambda}\in\sec T^{\ast}M\otimes T^{\ast}M\otimes TM. \label{8}%
\end{equation}

As can be easily verified from Eq.(\ref{3b}) it is the case that%
\begin{equation}
K_{\mu\nu\lambda}^{\cdot\cdot\cdot}=g_{\lambda\alpha}K_{\mu\nu\cdot}%
^{\cdot\cdot\alpha}=-K_{\mu\lambda\nu}^{\cdot\cdot\cdot} \label{9}%
\end{equation}

The validity of Eq.(\ref{9}) lead many authors to say the contortion tensor is
antisymmetric in the two last indices. However, it is necessary to observe
here that (parodying G\"{o}ckeler and Sch\"{u}cker \cite{gs}) the
anti-symmetry is a \textit{bastard} one, since we are comparing the components
of $\boldsymbol{K}$ that live on different spaces, namely\ $T^{\ast}M$ and
$TM$.

Moreover, writing
\begin{equation}
\boldsymbol{K}=K_{\mu\cdot\nu}^{\cdot\beta\cdot}\vartheta^{\mu}\otimes
\vartheta_{\beta}\otimes\boldsymbol{e}^{\nu}\in\sec T^{\ast}M\otimes TM\otimes
T^{\ast}M. \label{11}%
\end{equation}
where like in \cite{4}%

\begin{equation}
K_{\mu\cdot\nu}^{\cdot\beta\cdot}:=g_{\nu\lambda}g^{\beta\kappa}K_{\mu
\kappa\cdot}^{\cdot\cdot\lambda},\nonumber
\end{equation}
we have again a \textit{bastard} anti-symmetry since%
\begin{equation}
K_{\mu\cdot\nu}^{\cdot\beta\cdot}=-K_{\mu\nu\cdot}^{\cdot\cdot\beta}.
\label{11x}%
\end{equation}
\medskip

Finally we remark that (since Schouten \cite{3}) the $S_{\mu\nu\cdot}%
^{\cdot\cdot\lambda}=S_{\nu\mu\cdot}^{\cdot\cdot\lambda}$ are said to be the
components of the \emph{strain tensor}\textit{ }(of the connection $D$)%
\begin{equation}
\boldsymbol{S}=S_{\mu\nu\cdot}^{\cdot\cdot\lambda}\vartheta^{\mu}%
\otimes\vartheta^{\nu}\otimes\boldsymbol{e}_{\lambda}=S_{\mu\nu\lambda}%
^{\cdot\cdot\cdot}\vartheta^{\mu}\otimes\vartheta^{\nu}\otimes\boldsymbol{e}%
^{\lambda}\in\sec T^{\ast}M\otimes T^{\ast}M\otimes TM. \label{12}%
\end{equation}

\noindent\textbf{Remark} \emph{It is obvious from the above formulas that the
contorsion tensor is not antisymmetric in the lower indices }$\mu\nu$
\emph{due to the presence of the strain tensor that is symmetric contrary to
what is stated, e.g., in }\cite{5}.

\section{Proof of Eq.(\ref{3a})}

We start remembering that since $\mathring{D}\boldsymbol{g}=0$ and
$D\boldsymbol{g}=0$ we can write in an arbitrary coordinate basis that:%
\begin{align}
\mathring{D}_{\mu}g_{\nu\lambda}  &  =\partial_{\mu}g_{\nu\lambda}%
-\mathring{\Gamma}_{\mu\nu\cdot}^{\cdot\cdot\rho}g_{\rho\lambda}%
-\mathring{\Gamma}_{\mu\lambda\cdot}^{\cdot\cdot\rho}g_{\nu\rho}=0,\label{6}\\
D_{\mu}g_{\nu\lambda}  &  =\partial_{\mu}g_{\nu\lambda}-\Gamma_{\mu\nu\cdot
}^{\cdot\cdot\rho}g_{\rho\lambda}-\Gamma_{\mu\lambda\cdot}^{\cdot\cdot\rho
}g_{\nu\rho}=0. \label{6a}%
\end{align}

From Eq.(\ref{6}) and some trivial algebra we get, as well known%
\begin{equation}
\mathring{\Gamma}_{\mu\nu\cdot}^{\cdot\cdot\rho}=\frac{1}{2}g^{\lambda\rho
}\left(  \partial_{\mu}g_{\nu\lambda}+\partial_{\nu}g_{\mu\lambda}%
-\partial_{\lambda}g_{\mu\nu}\right)  . \label{7a}%
\end{equation}
From Eq.(\ref{6a}) we can write:%
\begin{align}
\partial_{\mu}g_{\nu\lambda}  &  =\Gamma_{\mu\nu\cdot}^{\cdot\cdot\rho}%
g_{\rho\lambda}+\Gamma_{\mu\lambda\cdot}^{\cdot\cdot\rho}g_{\nu\rho
,}\label{8a}\\
\partial_{\nu}g_{\mu\lambda}  &  =\Gamma_{\nu\mu\cdot}^{\cdot\cdot\rho}%
g_{\rho\lambda}+\Gamma_{\nu\lambda\cdot}^{\cdot\cdot\rho}g_{\mu\rho
,},\label{8b}\\
\partial_{\lambda}g_{\mu\nu}  &  =\Gamma_{\lambda\mu\cdot}^{\cdot\cdot\rho
}g_{\rho\nu}+\Gamma_{\lambda\nu\cdot}^{\cdot\cdot\rho}g_{\mu\rho}. \label{8c}%
\end{align}
Then,
\begin{align}
&  \partial_{\mu}g_{\nu\lambda}+\partial_{\nu}g_{\mu\lambda}-\partial
_{\lambda}g_{\mu\nu}\nonumber\\
&  =g_{\rho\lambda}(\Gamma_{\mu\nu\cdot}^{\cdot\cdot\rho}+\Gamma_{\nu\mu\cdot
}^{\cdot\cdot\rho})+g_{\nu\rho}(\Gamma_{\mu\lambda\cdot}^{\cdot\cdot\rho
}-\Gamma_{\lambda\mu\cdot}^{\cdot\cdot\rho})+g_{\mu\rho}(\Gamma_{\nu
\lambda\cdot}^{\cdot\cdot\rho}-\Gamma_{\lambda\nu\cdot}^{\cdot\cdot\rho}).
\label{9a}%
\end{align}

Observe that $\Gamma_{(\mu\nu)\cdot}^{\cdot\cdot\rho}:+\frac{1}{2}(\Gamma
_{\mu\nu\cdot}^{\cdot\cdot\rho}+\Gamma_{\nu\mu\cdot}^{\cdot\cdot\rho})$ is the
symmetric part of $\Gamma_{\mu\nu\cdot}^{\cdot\cdot\rho}$ whereas $\frac{1}%
{2}(\Gamma_{\mu\lambda\cdot}^{\cdot\cdot\rho}-\Gamma_{\lambda\mu\cdot}%
^{\cdot\cdot\rho})=\frac{1}{2}T_{\mu\lambda\cdot}^{\cdot\cdot\rho}$ is the
antisymmetric part of $\Gamma_{\mu\lambda\cdot}^{\cdot\cdot\rho}$. We can
rearrange the terms in Eq.(\ref{9a}) taking into account the definition of the
connections coefficients of the Levi-Civita connection $\mathring{D}$ as:%
\begin{align}
g_{\rho\lambda}\Gamma_{(\mu\nu)}^{\cdot\cdot\rho}  &  =\frac{1}{2}%
g_{\rho\lambda}(\Gamma_{\mu\nu\cdot}^{\cdot\cdot\rho}+\Gamma_{\nu\mu\cdot
}^{\cdot\cdot\rho})\nonumber\\
&  =\frac{1}{2}(\partial_{\mu}g_{\nu\lambda}+\partial_{\nu}g_{\mu\lambda
}-\partial_{\lambda}g_{\mu\nu})-\frac{1}{2}g_{\nu\rho}(\Gamma_{\mu\lambda
\cdot}^{\cdot\cdot\rho}-\Gamma_{\lambda\mu\cdot}^{\cdot\cdot\rho}%
)-g_{\rho\lambda}\frac{1}{2}(\Gamma_{\nu\lambda\cdot}^{\cdot\cdot\rho}%
-\Gamma_{\lambda\nu\cdot}^{\cdot\cdot\rho}). \label{10a}%
\end{align}

Then,
\begin{align}
\Gamma_{(\mu\nu)}^{\cdot\cdot\beta}  &  =\frac{1}{2}g^{\beta\lambda}%
(\partial_{\mu}g_{\nu\lambda}+\partial_{\nu}g_{\mu\lambda}-\partial_{\lambda
}g_{\mu\nu})\nonumber\\
&  -g^{\beta\lambda}\frac{1}{2}g_{\nu\rho}(\Gamma_{\mu\lambda\cdot}%
^{\cdot\cdot\rho}-\Gamma_{\lambda\mu\cdot}^{\cdot\cdot\rho})-g^{\beta\lambda
}g_{\rho\lambda}\frac{1}{2}(\Gamma_{\nu\lambda\cdot}^{\cdot\cdot\rho}%
-\Gamma_{\lambda\nu\cdot}^{\cdot\cdot\rho})\nonumber\\
&  =\mathring{\Gamma}_{\mu\nu\cdot}^{\cdot\cdot\beta}-\frac{1}{2}%
g^{\beta\lambda}(g_{\nu\rho}T_{\mu\lambda\cdot}^{\cdot\cdot\rho}%
+g^{\beta\lambda}g_{\rho\mu}T_{\nu\lambda\cdot}^{\cdot\cdot\rho})\nonumber\\
&  =\mathring{\Gamma}_{\mu\nu\cdot}^{\cdot\cdot\beta}+\frac{1}{2}S_{\mu
\nu\cdot}^{\cdot\cdot\beta} \label{10C}%
\end{align}
Finally, taking into account that $\Gamma_{\mu\nu\cdot}^{\cdot\cdot\rho
}=\Gamma_{(\mu\nu)\cdot}^{\cdot\cdot\rho}+\Gamma_{\lbrack\mu\nu]\cdot}%
^{\cdot\cdot\rho}=\Gamma_{(\mu\nu)\cdot}^{\cdot\cdot\rho}+\frac{1}{2}T_{\mu
\nu\cdot}^{\cdot\cdot\rho}$ we have \ using Eq.(\ref{10C})%
\begin{equation}
\Gamma_{\mu\nu\cdot}^{\cdot\cdot\rho}=\mathring{\Gamma}_{\mu\nu\cdot}%
^{\cdot\cdot\rho}+\frac{1}{2}S_{\mu\nu\cdot}^{\cdot\cdot\rho}+\frac{1}%
{2}T_{\mu\nu\cdot}^{\cdot\cdot\rho}=\mathring{\Gamma}_{\mu\nu\cdot}%
^{\cdot\cdot\rho}+K_{\mu\nu\cdot}^{\cdot\cdot\rho}, \label{10d}%
\end{equation}
and Eq.(\ref{3a}) is proved$\blacksquare$

\section{Relation Between the Curvature Tensors $\boldsymbol{R}$ and
$\boldsymbol{\mathring{R}}$}

Let $\boldsymbol{u,v,w\in}\sec TU$ and $\boldsymbol{\alpha\in}\sec%
{\textstyle\bigwedge\nolimits^{1}}
T^{\ast}U$. The curvature operators of $\mathring{D}$ and $D$ are defined by%

\begin{align}
\boldsymbol{\mathring{\rho},\rho}  &  :\sec TM\otimes TM\otimes
TM\longrightarrow\sec TM,\label{curvop}\\
\boldsymbol{\mathring{\rho}}(\boldsymbol{u},\boldsymbol{v},\boldsymbol{w})  &
=\mathring{D}_{\boldsymbol{u}}\mathring{D}_{\boldsymbol{v}}\boldsymbol{w}%
-\mathring{D}_{\boldsymbol{v}}\mathring{D}_{\boldsymbol{u}}\boldsymbol{w}%
-\mathring{D}_{[\boldsymbol{u,v}}]\boldsymbol{w},\\
\boldsymbol{\rho}(\boldsymbol{u},\boldsymbol{v},\boldsymbol{w})  &
=D_{\boldsymbol{u}}D_{\boldsymbol{v}}\boldsymbol{w}-D_{\boldsymbol{v}%
}D_{\boldsymbol{u}}\boldsymbol{w}-D_{[\boldsymbol{u,v}}]\boldsymbol{w.}%
\end{align}
It is usual to write \cite{choquet}\ $\boldsymbol{\mathring{\rho}%
}(\boldsymbol{u},\boldsymbol{v},\boldsymbol{w})=\boldsymbol{\mathring{\rho}%
}(\boldsymbol{u,v})\boldsymbol{w}$, $\boldsymbol{\rho}(\boldsymbol{u}%
,\boldsymbol{v},\boldsymbol{w})=\boldsymbol{\rho}(\boldsymbol{u,v}%
)\boldsymbol{w}$ and even to call curvature operators the objects%

\begin{align}
\boldsymbol{\mathring{\rho}}(\boldsymbol{u,v})  &  =\mathring{D}%
_{\boldsymbol{u}}\mathring{D}_{\boldsymbol{v}}-\mathring{D}_{\boldsymbol{v}%
}\mathring{D}_{\boldsymbol{u}}-\mathring{D}_{[\boldsymbol{u,v}}],\nonumber\\
\boldsymbol{\rho}(\boldsymbol{u},\boldsymbol{v})  &  =D_{\boldsymbol{u}%
}D_{\boldsymbol{v}}-D_{\boldsymbol{v}}D_{\boldsymbol{u}}-D_{[\boldsymbol{u,v}%
}]\boldsymbol{.} \label{13}%
\end{align}
Also, the Riemann curvature tensors of those connections are respectively the
objects:
\begin{align}
\boldsymbol{\mathring{R}}(\boldsymbol{w},\boldsymbol{u},\boldsymbol{v}%
,\boldsymbol{\alpha})  &  :=\boldsymbol{\alpha}%
(\boldsymbol{\boldsymbol{\mathring{\rho}}(u,v)w}),\nonumber\\
\boldsymbol{R}(\boldsymbol{w},\boldsymbol{u},\boldsymbol{v},\boldsymbol{\alpha
})  &  :=\boldsymbol{\alpha}(\boldsymbol{\boldsymbol{\rho}(u,v)w}). \label{14}%
\end{align}

Moreover, the components of the curvature tensors relative in the appropriated
coordinate basis associated to the coordinates $\langle x^{\mu}\rangle$
covering $U$ are:%
\begin{align}
\boldsymbol{\mathring{R}}(\boldsymbol{e}_{\mu},\boldsymbol{e}_{\alpha
},\boldsymbol{e}_{\beta},\boldsymbol{\vartheta}^{\lambda})  &  :=\mathring
{R}_{\mu\alpha\beta\cdot}^{\cdot\text{ }\cdot\text{ }\cdot\text{ }\lambda
}\nonumber\\
\boldsymbol{R}(\boldsymbol{e}_{\mu},\boldsymbol{e}_{\alpha},\boldsymbol{e}%
_{\beta},\boldsymbol{\vartheta}^{\lambda})  &  :=R_{\mu\alpha\beta\cdot
}^{\cdot\text{ }\cdot\text{ }\cdot\text{ }\lambda}. \label{15}%
\end{align}
We get after some trivial (but tedious algebra as in the last section)
\cite{2} that
\begin{equation}
R_{\mu\alpha\beta\cdot}^{\cdot\text{ }\cdot\text{ }\cdot\text{ }\lambda
}=\mathring{R}_{\mu\alpha\beta\cdot}^{\cdot\text{ }\cdot\text{ }\cdot\text{
}\lambda}+J_{\mu\lbrack\alpha\beta]\cdot}^{\cdot\text{ \ }\cdot\text{ \ }%
\cdot\text{ \ }\lambda} \label{16}%
\end{equation}
where
\begin{align}
J_{\mu\alpha\beta\cdot}^{\cdot\cdot\cdot\lambda}  &  =D_{\alpha}K_{\beta
\mu\cdot}^{\text{ }\cdot\cdot\lambda}-K_{\beta\sigma\cdot}^{\cdot\cdot\lambda
}K_{\alpha\mu\cdot}^{\cdot\cdot\sigma}\nonumber\\
J_{\mu\lbrack\alpha\beta]\text{ }\cdot}^{\cdot\text{ \ }\cdot\text{ \ }%
\cdot\text{ \ }\lambda}  &  =J_{\mu\alpha\beta\text{ }\cdot}^{\cdot\text{
}\cdot\text{ }\cdot\text{ }\lambda}-J_{\mu\beta\alpha\text{ }\cdot}%
^{\cdot\text{ }\cdot\text{ }\cdot\text{ }\lambda}. \label{17}%
\end{align}

\section{Geometry of a Manifold where $\Gamma_{\mu\lambda\cdot}^{\cdot
\cdot\rho}=-\Gamma_{\lambda\mu\cdot}^{\cdot\cdot\rho}$ in some Coordinates
$\langle x^{\mu}\rangle$ Covering $U\subset M$}

Does the condition $\Gamma_{\mu\lambda\cdot}^{\cdot\cdot\rho}=-\Gamma
_{\lambda\mu\cdot}^{\cdot\cdot\rho}$ implies that $\mathring{\Gamma}%
_{\mu\lambda\cdot}^{\cdot\cdot\rho}=0$? Of course, \textit{not in general}.
Let us see the reason for that. From Eq.(\ref{3a}) we see that all that is
necessary for the validity of $\Gamma_{\mu\lambda\cdot}^{\cdot\cdot\rho
}=-\Gamma_{\lambda\mu\cdot}^{\cdot\cdot\rho}$ is that in the coordinate system
where $\Gamma_{\mu\lambda\cdot}^{\cdot\cdot\rho}=-\Gamma_{\lambda\mu\cdot
}^{\cdot\cdot\rho}$ we have
\begin{equation}
\mathring{\Gamma}_{\mu\lambda\cdot}^{\cdot\cdot\rho}=-\frac{1}{2}S_{\mu
\lambda\cdot}^{\cdot\cdot\rho} \label{18}%
\end{equation}

Observe moreover that when $\Gamma_{\mu\lambda\cdot}^{\cdot\cdot\rho}%
=-\Gamma_{\lambda\mu\cdot}^{\cdot\cdot\rho}$ and besides that we have also
$S_{\mu\lambda\cdot}^{\cdot\cdot\rho}=0$, then it follows from Eqs.(8) that
$\partial_{\mu}g_{\nu\lambda}=0$, i.e., the $g_{\nu\lambda}$ are
\textit{constant functions} of the coordinates. Then, taking into account the
Lorentz signature of the metric we can introduce coordinates \ in some open
set intersecting $U$ such that the matrix with entries $g_{\nu\lambda}%
^{\prime}$ is the diagonal matrix \textrm{diag}$(1,-1,-1,-1)$.

Thus \ we see that \ taking into account Eq.(\ref{10C}) \ when $\Gamma
_{\mu\lambda\cdot}^{\cdot\cdot\rho}=-\Gamma_{\lambda\mu\cdot}^{\cdot\cdot\rho
}$ we have that $\mathring{\Gamma}_{\mu\lambda\cdot}^{\cdot\cdot\rho}=0$ only
if $S_{\mu\lambda\cdot}^{\cdot\cdot\rho}=0$.

However, it is a good idea to keep in mind that even in that case the Riemann
curvature tensor of the connection$\ $Riemann-Cartan connection $D$ is not
null in general. Indeed, from Eqs.(\ref{16}) and (\ref{17}) it follows that
\begin{align}
R_{\mu\alpha\beta\cdot}^{\cdot\text{ }\cdot\text{ }\cdot\text{ }\lambda}  &
=J_{\mu\lbrack\alpha\beta]\text{ }\cdot}^{\cdot\text{ \ }\cdot\text{ \ }%
\cdot\text{ \ }\lambda}\nonumber\\
&  =D_{\alpha}K_{\beta\mu\cdot}^{\text{ }\cdot\cdot\lambda}-K_{\beta
\sigma\cdot}^{\cdot\cdot\lambda}K_{\alpha\mu\cdot}^{\cdot\cdot\sigma}%
-D_{\beta}K_{\alpha\mu\cdot}^{\text{ }\cdot\cdot\lambda}+K_{\alpha\sigma\cdot
}^{\cdot\cdot\lambda}K_{\beta\mu\cdot}^{\cdot\cdot\sigma} \label{19}%
\end{align}

Finally, we ask: what is a sufficient condition for $R_{\mu\alpha\beta\cdot
}^{\cdot\text{ }\cdot\text{ }\cdot\text{ }\lambda}=0$?

The condition is the existence of four \textit{parallel vector fields} defined
on all $M$ such that they are basis for each $T_{x}M$ (the tangent space at
$x\in M$). A set of parallel vector fields $\{\boldsymbol{X}_{a}\}$,
$a=0,1,2,3$ is by definition one such that $D_{\boldsymbol{X}_{a}%
}\boldsymbol{X}_{b}=0$ for all $\ a,b=0,1,2,3$. A space with this property is
called parallelizable and in the case where it has Lorentzian metric is known
as \emph{Weintzbock spacetime}.

\section{Final Remarks}

In his note 122 \cite{6} MWE states as `theorem' that \emph{any} connection
must be anti-symmetric. From Eq.(\ref{3b}) above it is obvious that this
statement is simply $\boldsymbol{wrong}$. That error simply invalidates almost
all of his statements presented in his series of papers on `ECE theory'. And
indeed, it is well known that some (if not all) of those papers are full of
very serious errors, including one that MWE calls the `\emph{dual Bianchi
identity}', a non sequitur that leads him to claim that Einstein's equations
are mathematically wrong! Of course they are not. For more details on this
particular issue see \cite{7}. For a discussion of some another MWE serious
flaws see also \cite{3,9,10,11}.

Those sad facts are being presented here because despite the criticisms quoted
above that simply show that ECE theory is a nonsequitur, MWE recently found
support from a british publisher\footnote{See:
http://www.cisp-publishing.com/} to launch a journal: \emph{Journal of
Foundations of Physics and Chemistry} which will publish all the papers he and
others authored on \textit{ECE}. Before you order such a journal, please give
a read with attention on this pedagogical and \emph{free} note and also the
\emph{free} references quoted below.

\end{document}